\documentclass[manuscript]{aastex}
\usepackage{amssymb}
\usepackage{pifont}
\usepackage{graphicx}
\usepackage{multirow}

\shortauthors{Yan et al.}
\begin{document}

\title{Time evolution of the X-ray and $\gamma$-ray fluxes of the Crab pulsar}
\author{L. L. Yan\altaffilmark{1,2,3}, M. Y. Ge\altaffilmark{2}, F. J. Lu\altaffilmark{2}, S. J. Zheng\altaffilmark{2},
Y. L. Tuo\altaffilmark{2,3}, Z. J. Li\altaffilmark{2,3}, L. M. Song\altaffilmark{2}, J. L. Qu\altaffilmark{2}}
\affil{$^1$School of Mathematics and Physics, Anhui Jianzhu University, Hefei 230601, China;}
\affil{$^2$Key Laboratory of Particle Astrophysics, Institute of High Energy Physics,
Chinese Academy of Sciences, Beijing 100049, China; yanlinli@ihep.ac.cn; gemy@ihep.ac.cn; lufj@ihep.ac.cn}
\affil{$^3$University of Chinese Academy of Sciences, Beijing 100049, China.}

\begin{abstract}
We studied the evolution of the X-ray and $\gamma$-ray spectra of the Crab pulsar utilizing
the 11-year observations from the Rossi X-ray Timing Explorer ({\sl{RXTE}}) and 9-year
observations from the Fermi Gamma-ray Space Telescope ({\sl{FGST}}). By fitting the spectrum
of each observation, we obtained the corresponding flux, and then analysed the long term
evolution of the X-ray (or $\gamma$-ray) luminosities as well as their correlations with the
spin down power of the pulsar. The X-ray flux in 5--60\,keV obtained by the Proportional Counter
Array (PCA) of {\sl{RXTE}} decreases with a rate of $(-2.4\pm0.4)\times 10^{-14}$ erg cm$^{-2}$ s$^{-1}$
per day. The X-ray flux in 15-250\,keV obtained by the High Energy X-ray Timing Experiment (HEXTE)
of {\sl{RXTE}} and the $\gamma$-ray flux in 0.1--300\,GeV by the Large Area Telescope (LAT) onboard
{\sl{FGST}} show similar decreasing trend, but are unsignificant statistically. The 5--60\,keV X-ray
luminosity $L_{X}$ is correlated with the spin down power $L_{sd}$ by $L_{X}{\propto}L_{sd}^{1.6\pm0.3}$,
which is similar to the statistical results for young pulsars.
\end{abstract}


\section{Introduction}

Much effort has been devoted to statistical studies of pulsar high energy
emission properties, with particular emphasis on the efficiency of
the conversion of the pulsar spin down power $L_{sd}$ into X-ray and $\gamma$-ray luminosities,
so as to test the pulsar emission models and to predict the properties of a specific source.
Previous studies show the X-ray luminosities $L_{X}$
of pulsars have a correlation with their spin down power $L_{sd}$. \cite{Seward1988}
first found that the 0.2--4\,keV X-ray luminosity $L_{X}{\propto}L_{sd}^{1.39}$.
\cite{BT1997} used 27 pulsars detected in 0.1--2.4\,keV and found that
$L_{X}\propto{L_{sd}}$. In the higher energy band, \cite{Saito1998} gave the relation
$L_{X}\propto{L_{sd}^{1.5}}$ in 2--10\,keV, but \cite{Possenti2002} obtained a relation
of $L_{X}{\propto}L_{sd}^{1.34}$ in the same energy band using a larger sample of 39 pulsars.
However, all the X-ray luminosities above contain the emission from the pulsar wind nebula (PWN),
which is nonpulsed and often dominates the nonthermal emission of pulsars. To eliminate the
influence of the PWN emission, \cite{Cheng1999} presented the empirical relation between the
pulsed luminosity and the spin down power, i.e. $L_{X}{\propto}L_{sd}^{1.15}$ in 2--10\,keV.
Because of the improvement of spatial resultion of detectors,
\cite{Cheng2004} divided the total X-ray emission into pulsed and nonpulsed components,
and found that the pulsed component follows $L_{X}{\propto}L_{sd}^{1.2\pm0.08}$
in 2--10\,keV, which is more gentle than the nonpulsed component and can be
explained by the outer gap radiation model \citep{ZJ2006}. By using the {\sl Chandra}
and {\sl XMM-Newton} observations, \cite{Li2008} resolved the pulsar X-ray emission
from that of the PWNe and got $L_{X}{\propto}L_{sd}^{0.92\pm0.04}$ in
the same energy band. The relation between the pulsar $\gamma$-ray luminosity $L_{\gamma}$
and the spin down luminosity are different from that in the X-ray band. \cite{Saito1997}
obtained $L_{\gamma}{\propto}L_{sd}^{0.5}$, and \cite{Marelli2011} further found that the
relation between ${\rm log}_{10}$($\,L_{\gamma}$) and ${\rm log}_{10}$($L_{sd}$) could not
be represented by a single linear function. These correlations are in accord with the
expectation that the X-ray and $\gamma$-ray emission of a pulsar is at the expense of rotational
energy. This could be from magnetosphere and near the light cylinder \citep{BT1997, Malov2003}.
Specific predictions of the correlations for emission from outside the light cylinder have not
yet been given.

However, all the previous results are obtained from the statistics of pulsar samples. The
different properties of these pulsars, such as ages, magnetic field strengths, masses, and
directions of the magnetic and rotation axes, bring complexity to these relations, which may
in turn make it more difficult to understand the physics process behind. Therefore, to obtain
the exact dependence of the high energy luminosity on the evolutionary spin down power of a
specific pulsar in a long time span could provide much purer information for pulsar physics study.

Among all the pulsars, the Crab pulsar is the most suitable source for such studies, because it
has been frequently and comprehensively studied in almost all wavelength bands from radio to very
high energy $\gamma$-rays. Recently, for this pulsar, secular changes with X-ray pulse profiles
and phases are reported \citep{Ge et al.(2012), Ge et al.(2016), Yan et al.(2017)}. But,
the evolution of the X-ray and $\gamma$-ray spectra have not been studied in details yet, which
are what we want to examine in the current paper. Thanks to the long term observation from the
{\sl {RXTE}} and the {\sl FGST}, we can study the evolution of the spectra in 5--60\,keV, 15--250\,keV
and 0.1--300\,GeV, using the PCA and the HEXTE of {\sl {RXTE}} and the LAT onboard {\sl FGST}, respectively.
These results are then used to test the pulsar emission models.

The organization of this paper is as follows: the data processing and reduction are presented in
Section 2, results are given in Sections 3, discussions on the physical implications of our results
in Section 4, and a short summary in Section 5. All through the paper, the errors of the parameters
are at 1\,$\sigma$ level.

\section{Observations and Data Processing}

\subsection{Timing Ephemeris from Jodrell Bank Observatory}

A 13-m radio telescope at Jodrell Bank Observatory (JBO) monitors
the Crab pulsar daily \citep{Lyne(1993)}, offering a radio ephemeris \footnote{http://www.jb.man.ac.uk/pulsar/crab.html}
that is used for the analyses of {\sl RXTE} and {\sl FGST} data.
For monthly validity intervals, this ephemeris contains up-to-date
parameters of the rotation frequency and its first two time
derivatives. The radio pulsar position is R.A.=05h\,34m\,31.972s,
decl.=+22$^\circ$\,00$^\prime$\,52.07$^{\prime\prime}$ (J2000). These
parameters are essential in the phase folding step of data analysis.
Besides the pulse profile folding, we also use this ephemeris to reckon
the spin period and its derivative at a certain time.

\subsection{{\sl RXTE} Observations and Spectra Fitting Method}

The X-ray data used in this work were obtained by both PCA and HEXTE, which are
described in details in \cite{Rothschild et al.(1998)}, \cite{Jahoda et al.(2006)}
and \cite{Yan et al.(2017)}. The PCA data were taken in MJD 51983--55928 (UTC 2000
December 15--2012 January 02) and are in event mode E\_250us\_128M\_0\_1s, while
the HEXTE data were taken in MJD 51302--55913(UTC 1999 May 04--2011 December 18) and
are in data mode E\_8us\_256\_DX0F. However, even in these durations, observations
with short exposure time and thus limited photon counts were not used, to ensure small
statistic errors in our analyses.

The {\sl {RXTE}} data  were analyzed by using the FTOOLS from the astronomy software
HEASOFT (v6.17). To get the spectrum and flux from each observation, we first select
the good events and then fold them by using the FTOOLS commands {\sl fasebin} and {\sl fbssum}.
As we focus on the X-ray properties of the Crab pulsar, we need to subtract the background
from the nebula. To do this, we use the command {\sl fbfsum} to get the spectra in the
whole phase range and that in phase 0.6--0.8 (off-pulse window) which is similar to that in \cite{Ge et al.(2016)},
and the later is taken as the contribution from the nebula.

The XSPEC in the software package HEASOFT is used to fit the spectrum of each observation by
using the files produced above. The fitting model we used is {\sl{wabs}} combined with {\sl{powerlaw}}.
The first component is used to represent the absorption of the interstellar medium, whose column
density $N_{H}$ is fixed as $0.36\times10^{22}$ cm$^{-2}$ \citep{Ge et al.(2012)}, and the second
component is used to fit the shape of the spectrum of the pulsar X-ray emission. By fitting the PCA
and HEXTE spectra of each observation, we got the photon indices and fluxes of the pulsar in 5--60\,keV
and 15--250\,keV, and these values are further averaged over a duration of about 100 days and 400 days,
respectively, as shown in Fig. \ref{fig1}.

\subsection{{\sl FGST} Observations and Spectra Fitting Methond}

The purpose of {\sl FGST}/LAT is to detect $\gamma$-rays in the energy range from 20 MeV
to 300\,GeV, with an effective area of $\sim$ 8000\,cm$^2$. It consists of a high-resolution
converter tracker, a CsI(Tl) crystal calorimeter, and an anti-coincidence detector, which
could measure the direction and energy of the $\gamma$-rays, and in the same time discriminate
the particle background events \citep{Atwood2009}. The $\gamma$-ray events from MJD 54771--57966
(UTC 2008 November 01--2017 August 01) in 0.1--300\,GeV are analysed in this paper. To balance the
requirements of statistics and evolution study, the observations are divided into 15 groups, and each group
has a time span of about 230 days.

The LAT data analysis was accomplished by using the Fermi Science Tools (v10r0p5). First, we
use the commands {\sl gtselect} and {\sl gtmktime} to select the events in good time intervals.
Then, in order to reduce the contamination from neighboring sources and to get more accurate spectral
fitting results, photon events in the region of 10$^\circ$ around the pulsar radio position are
selected, which is smaller than previously used \citep{Abdo et al.(2010)}.
At the end, the timing ephemeris from JBO was used to calculate the phase of each $\gamma$-ray photon, and photons in phase
0.52--0.87 (off-pulse, \cite{Abdo et al.(2010)}) are used for nebula spectrum fitting.

The analysis of $\gamma$-ray spectra was performed by
using a maximum-likelihood method implemented in the Fermi science
analysis tools
\footnote{https://fermi.gsfc.nasa.gov/ssc/data/analysis/scitools/binned\_likelihood\_tutorial.html}.
The events within 10$^\circ$ from the Crab pulsar contain the emission from the Crab pulsar itself, the Crab nebula, the nearby sources, and
the diffuse $\gamma$-ray background. To get the spectral parameters of the pulsar reliably, we
determine the spectral parameters of the other three components first, and then use these
information in the final spectral fitting. The spectral
parameters of the
Crab nebula were obtained by using the photons in phase 0.52--0.87
that were selected previously.
 Contribution of all the neighboring sources (within 15$^\circ$) are considered, in which the spectral parameters of sources
more than 3$^\circ$ away from the pulsar direction were fixed as those obtained from the all-sky analysis, and those for sources
within 3$^\circ$ were taken as free parameters.
The contamination produced by the interactions between the cosmic-rays
and the Earth's atmosphere was avoided by selecting
a zenith angle greater than $105^\circ$. The Galactic diffuse emission model we used is gll\_iem\_v06.fits, the isotropic model
is iso\_P8R2\_SOURCE\_V6\_v06.txt, LAT 4-year Point Source
Catalog is gll\_psc\_v16.fit \citep{Acero2015ApJS}, and the instrument response function is P8R2\_SOURCE\_V6. At the end, the
spectra of the Crab pulsar were obtained by utilizing events in the whole phase range with the contaminations from the other components
taken into account. The detailed spectral analysis process is the same
as in \cite{Abdo et al.(2010)}.

From the spectral analysis results, we can compute the X-ray and $\gamma$-ray luminosities
$L_{X,\gamma}=4{\pi}d^2f_{X,\gamma}F_{X,\gamma}$ \citep{Marelli2011} and then study their
correlation with the spin down power $L_{sd}=\frac{4\pi^2I\dot{P}}{{P}^3}$ \citep{Condon2016book},
where $d=2$\,kpc \citep{Trimble1973} is the pulsar distance, $F_{X}$ and $F_{\gamma}$ are the
X-ray and $\gamma$-ray fluxes, $f_{X}$ and $f_{\gamma}$ account for the X-ray and $\gamma$-ray
beaming geometries that depend on the viewing angle and the magnetic inclination of the pulsar,
which are taken as 1 \citep{Marelli2011}, $P$ and $\dot{P}$ are the spin period and its derivative,
$I=\frac{2}{5}MR^2$ is the neutron star momentum, $M\simeq1.4M_{\odot}$ is the mass of the Crab
pulsar, and $R\simeq10$\,km is the typical radius of a neutron star.

\subsection{Linear Fitting}

The long term evolutions of the fluxes and photon indices are studied by a linear fitting
method that was used by \cite{Li2008} and \cite{Ge et al.(2012)}, and the relation between
the high energy luminosities and the spin down power are studied in a similar way. The value
of the slope and its error obtained from the fitting show whether there are any long term
evolution and dependence. To double check, we also use robust linear modeling (RLM) to fit
the evolution and dependance trends.

The RLM is inherited from the R statistical software package \citep{Feigelson(2012)}.
The \emph{MASS} (Modern Applied Statistics with S) library based on R-language has
the \emph{rlm} function for RLM, in which the best fit is achieved using an iteratively
reweighted least-squares algorithm. This method is used to fit the spectral parameters
versus time and the correlation between different data groups. The fitting results are
listed in Table \ref{table:1} and Table \ref{table:2}.

\subsection{Correlation Analysis}
The Pearson correlation coefficient is a suitable parameter to describe the dependance of
a parameter on the other parameters quantitatively \citep{Lee Rodgers(1988)}. In order to
evaluate the significance level of the correlations of two parameters concerned, we evaluate
the widely used Pearson correlation coefficient ($\rho$) and the two-sided significance level
($p$). The results are listed in Table \ref{table:2}.

\section{Results}

\subsection{The Spectra and Energy Conversion Efficiency Evolution}
\begin{figure}
\begin{center}
\includegraphics[width=1.0\textwidth,height=0.7\textwidth]{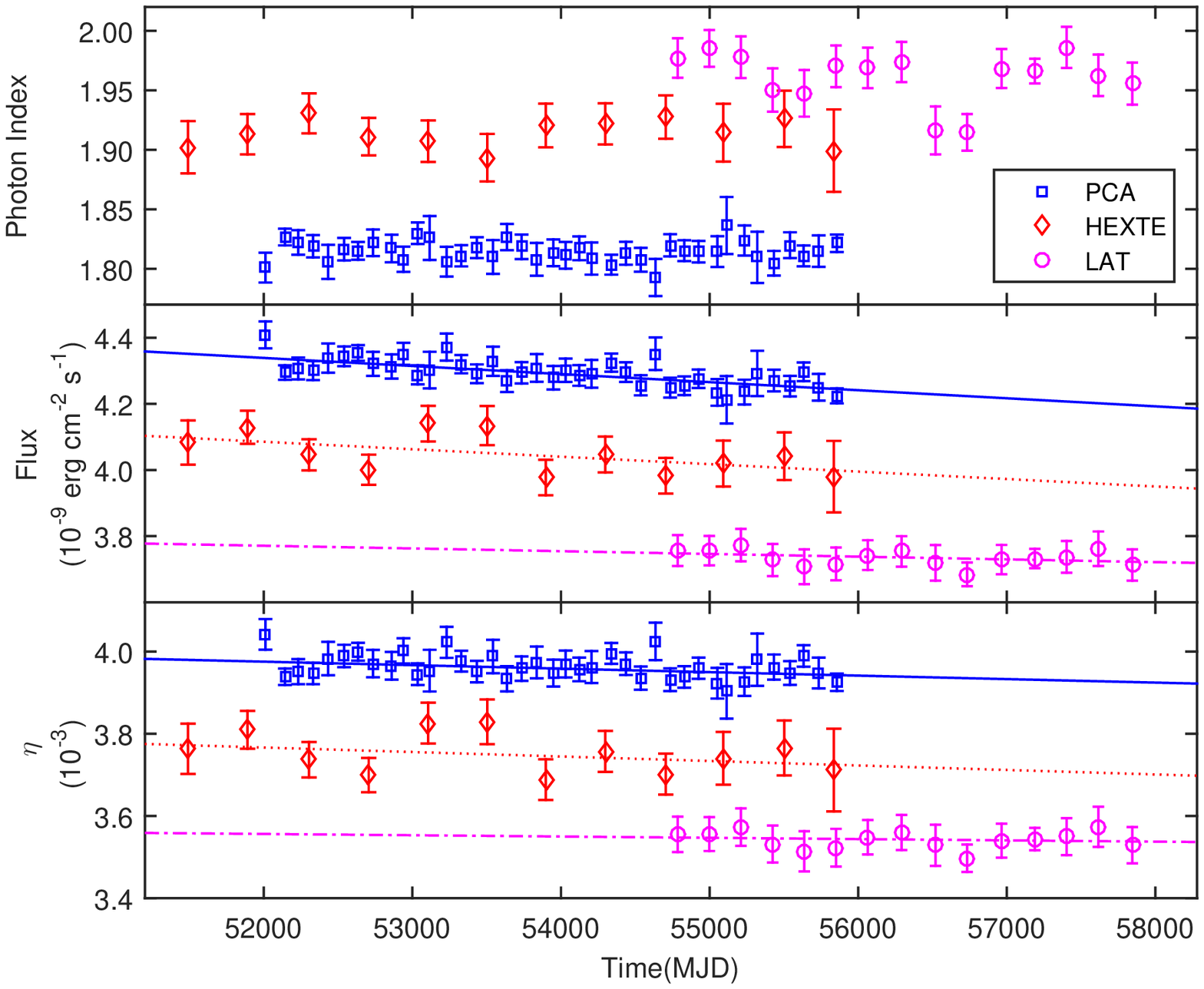}
\caption{The spectra and energy conversion efficiency evolutions of the Crab pulsar in X-ray
and $\gamma$-ray. In order to compare these results clearly, the data points of the flux from HEXTE
and LAT are shifted upward about $0.9\times10^{-9}$, $2.4\times10^{-9}$\,erg\,cm$^{-2}$\,s$^{-1}$,
respectively, and the data points of conversion efficiency $\eta$ from HEXTE and LAT are shifted upward
about $0.85\times10^{-3}$, $2.3\times10^{-3}$, respectively. The blue solid line, red dotted line, and
carmine dotted¨Cdashed line are the fitting results for the data points from PCA, HEXTE, LAT, respectively.
\label{fig1}}
\end{center}
\end{figure}

Given that the rotation parameters of the Crab pulsar are evolving with time, and the
previous statistical studies show strong correlation between the high energy luminosities
and the spin down power of isolated pulsars, what we are interested in is whether the observed
high energy emission properties of the Crab pulsar change with time. In the upper two panels
of Fig. \ref{fig1}, the photon indices and fluxes obtained from the three instruments are plotted.
The photon indices don't have any obvious evolution with time, and the mean spectral indices from
PCA, HEXTE and LAT are 1.81, 1.91, 1.96, respectively, these results are consistent with previous
studies \citep{Kuiper2001, Abdo et al.(2010)}. The X-ray and $\gamma$-ray fluxes show a
similar decrease trend with time. Linear fitting to these data points gives that, the PCA flux
decreases with a rate of $(-2.4\pm0.4)\times10^{-14}$ erg\,cm$^{-2}$\,s$^{-1}$ per day, the HEXTE
flux decreases with a rate of $(-2.3\pm1.8)\times10^{-14}$\,erg\,cm$^{-2}$\,s$^{-1}$ per day, and the LAT flux decreases with
$(-0.8\pm0.7)\times10^{-14}$ erg\,cm$^{-2}$\,s$^{-1}$ per day. The errors listed here are the statistical
errors only. However, the instrument effect for the observations from PCA and HEXTE are very small
and could be neglected \citep{Ge et al.(2016)}. So, the PCA flux has decreased significantly in the 11
years of observation and this trend is consistent with previous studies in \cite{Wilson2011ApJ}, which show that
the Crab pulsar's flux steadily decreases at $\sim$0.2\% per year,
while the evolution of the HEXTE and LAT fluxes have not been detected significantly,
probably due to the limited number of photons detected.

The bottom panel of Fig. \ref{fig1} plots the conversion efficiency versus time in each energy band,
which is simply the ratio of the X-ray or $\gamma$-ray luminosity to the spin down power of the pulsar,
$\eta_{X,\gamma}=\frac{L_{X,\gamma}}{L_{sd}}$. The mean energy conversion efficiencies in the three
energy bands are 0.40\%, 0.29\% and 0.12\%, respectively. The $\eta_{X,\gamma}$ in this work is consistent with the constraints for young pulsars in \cite{Possenti2002}.

The energy conversion efficiency in 5--60\,keV
decreases with time, with a rate of $(-8.5\pm3.8)\times10^{-9}$ per day. But the conversion efficiencies
in the higher energy bands inferred from the HEXTE and LAT data don't show any detectable evolution, due
to the limited number of photons detected again.

\subsection{The Correlation between the X-ray Luminosity and Spin down Power}

\begin{figure}
\begin{center}
\includegraphics[width=1.0\textwidth,height=0.75\textwidth]{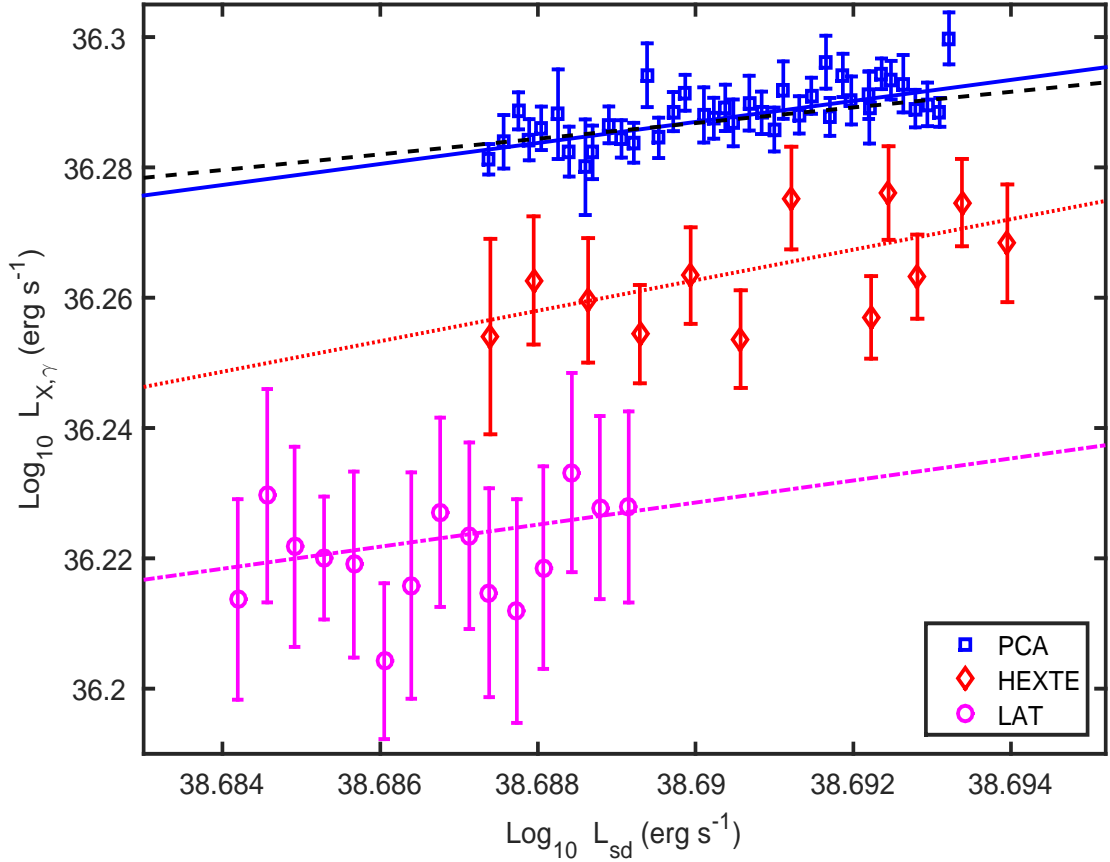}
\caption{The high energy luminosities vs. spin down powers. In order to compare these results
clearly, the data points from HEXTE and LAT are shifted upward about $0.11$, $0.44$, respectively.
The blue solid line, red dotted line, and carmine dotted¨Cdashed line are the fitting results for the
data points from PCA, HEXTE, LAT, respectively. The black dashed line is the correlation trend between
total theoretical X-ray luminosities and spin down powers based on the outer gap model in \cite{Cheng1998ApJ},
which is shifted upward 0.589 to compared with data points from PCA clearly.
\label{fig2}}
\end{center}
\end{figure}

As shown in Fig. \ref{fig2}, there exist clear positive correlations between the observed
high energy luminosity and the spin down power. The linear fitting parameters of the
$L_{X,\gamma}$--${L_{sd}}$ relations are listed in Table \ref{table:1}, in which the
Pearson correlation tests are: $\rho=0.69$ and significance level $p<10^{-5}$ for the
observations by PCA, $\rho=0.59$ and $p=0.05$ for the HEXTE observations, and $\rho=0.30$
and $p=0.28$ for the LAT observations. Therefore,the correlation between
$L_X$ and $L_{sd}$ is significantly detected, while that between $L_{\gamma}$ and $L_{sd}$
is undetected with the LAT data.

We further fit the relations between $L_{X,\gamma}$ and $L_{sd}$ with the function
$L_{X,\gamma} = n\,L_{sd}^m$, where $m$ and $n$ are the free parameters need to be
determined. This function is the same as used in the previous statistical studies
\citep{BT1997, Saito1998, Cheng2004, Li2008}. From the PCA data, we obtained a relationship
$L_{X}{\propto}L_{sd}^{1.6\pm0.3}$. For the HEXTE and LAT data, the relations are
$L_{X}{\propto}L_{sd}^{2.3\pm1.5}$, and $L_{\gamma}{\propto}L_{sd}^{1.7\pm1.4}$, which
means that the indices have not been well determined.

\section{Discussions}

\subsection{Comparisons with the Previous Works}
In this work, we find that the pulsed X-ray luminosity of the Crab pulsar
decreases significantly in the 11 years of {\sl RXTE}/PCA observations,
and it is strongly correlated with the spin down power by $L_{X}{\propto}L_{sd}^{1.6\pm0.3}$.
This is well consistent with the value found by \cite{Vink2011ApJ} in a study of
young pulsars, but a little higher than the values obtained in older statistical studies, of
about 1.0 \citep{Cheng2004, Li2008}.
We suggest that this discrepancy is probably due to the different samples used in these studies.
As dozens of pulsars were used in the previous studies, and the magnetic field strength,
inclination angle of the magnetic axis, rotation frequency, direction of the
rotation axis, and age are different for one pulsar from the others, the
previous statistical results are actually the combined effects of all these
factors rather than the consequence of the spin down power alone. Our results
thus provide relatively purified data to study the changes of the structure and
energy release mechanism of the pulsar magnetosphere when the spin down power decreases.

\subsection{Constraints on Pulsar Radiation Models}

There are different radiation and magnetosphere models to explain the high energy emission properties of pulsars, such as
vacuum magneto-dipole radiation model \citep{Pacini1968}, longitudinal current model \citep{Beskin1983}, outer gap
model (hereafter OG model, \cite{Cheng1986ApJ}) and FIDO (Force-free Inside, Dissipative Outside) model \citep{Spitkovsky2006ApJ}, and the relation
between high energy luminosities and spin down powers predicted by different models could be a good tool to test these models.
It is notable that the OG model in \cite{Cheng1998ApJ}, which used the thick outer gap model of \cite{Zhang1997ApJ} gave a prediction for the relation between X-ray ($<$1\,MeV) luminosities and spin down powers,
\begin{equation}
L_{sd}^{\rm (OG)}=3.8\times10^{31}{B_{0}(\rm G)^2}P(\rm s)^{-4} {\rm erg\,s^{-1}},
\label{eq4}
\end{equation}

\begin{equation}
L_{X,th}^{\rm (OG)}=5.5\times10^{-4}(\frac{\tan{\alpha}}{\tan{55^\circ}})^4{B_{0}}^{0.13}_{12}P^{-0.8}_{-1}L_{sd}^{\rm(OG)}.
\label{eq5}
\end{equation}
where $\alpha$ is the inclination angle, $B_{0}$ is the surface magnetic field strength, $L_{sd}^{\rm (OG)}$ is the spin down power and
$L_{X,th}^{\rm (OG)}$ is the theoretical prediction for total X-ray luminosity.
In the following we test this theoretical relation of X-ray luminosity and spin down power by comparing it with our observational results.
We simply take $B_{0}=10^{12.58}$\,G \citep{Cheng1986ApJ} and
$\alpha=45^\circ$ \citep{Du2012} because the correlation exponent $m$ for the OG model dose not change for different $B_{0}$ and $\alpha$.
As shown in Fig. \ref{fig2}, the correlation between the total theoretical X-ray luminosities and spin down powers has the same trend with
observations in selected energy bands (i.e. PCA observations). The energy bands of the theoretical predictions and observations are different,
so the normalization parameter $n$ of them are not the same. The relation between the theoretical X-ray luminosities and spin down powers is
$L_{X,th}^{\rm (OG)}{\propto}{L_{sd}^{\rm (OG)}}^{1.2}$, and the correlation exponent $m$ predicted by OG model is in the 1.5$\sigma$ limits
of the observational index. Thus the OG model we used is compatible with the 5--60\,keV observations, but as pointed out by \cite{Vigano2015}
the model as well as the data has significant uncertainties.

Alternatively, we realised that the decrease of the X-ray luminosity could be also due to the geometrical
evolution of the magnetosphere. Studies of the radio and X-ray profiles show that the separation and
relative strength of the two peaks changes gradually with time, which were explained as the increase of
the inclination angle of the magnetic axis \citep{Lyne2013Sci, Ge et al.(2016)}. Possibly,
the evolution of the inclination angle also changes the geometry of the X-ray beam and consequently makes
the X-ray flux lower.

\section{Summaries}

In this work, using the available observations of the {\sl RXTE}/PCA, {\sl RXTE}/HEXTE
and {\sl FGST}/LAT in 5--60\,keV, 15--250\,keV and 0.1--300\,GeV, we obtain the spectral
properties of the Crab pulsar in these energy bands. We find that the 5-60\,keV X-ray
flux observed by PCA has significantly decreased in the 11 years with a rate of
$(-2.4\pm0.4)\times10^{-14}$\,erg\,cm$^{-2}$\,s$^{-1}$ per day, while in the higher energy
bands this trend is not significant, probably due to the limited number of photons collected.
We also find a strong correlation between the 5-60\,keV X-ray luminosities and the spin down powers
of the Crab pulsar, which can be represented by $L_{X}{\propto}L_{sd}^{1.6\pm0.3}$.
Such a correlation is consistent with the predictions of the OG model in \cite{Cheng1998ApJ},
but may also be consistent with other models.

\section*{Acknowledgments}
We thank the referee for his/her very helpful comments. We appreciate Jian Li, Feifei Kou, Xian Hou, and Shanshan Weng for their useful suggestions and discussions. We thank the
High Energy Astrophysics Science Archive Research Center and Fermi Science Support Center
for maintaining its online archive service that provided the data used in this
research. This work is supported by the National Key R\&D Program of China (2016YFA0400803, 2016YFA0400800), National
Science Foundation of China (11233001, 11503027 and 61471003), the Strategic Pionner Program on Space Science,
and Chinese Academy of Sciences, Grant No.XDA15051400.

\clearpage

\clearpage

\begin{table}
\footnotesize
\caption {The evolution of spectra and conversion efficiency for the Crab pulsar$^{[a]}$}
\scriptsize{}
\label{table:1}
\medskip
\begin{center}
\begin{tabular}{c c c c c c c c}

\hline \hline
\multirow{2}{*}{Instrument} & \multirow{2}{*}{Energy} & \multicolumn{2}{c}{Flux (erg cm$^{-2}$ s$^{-1}$)} & & \multicolumn{2}{c}{$\eta$}  \\
 \cline{3-4} \cline{6-7}
   &    & Rate ($10^{-14}$/day) & Intercept (10$^{-9}$ ) & &Rate ($10^{-9}$/day) & Intercept (10$^{-3}$ )   \\
\hline
   PCA      &    5--60\,keV    & $-2.4\pm0.4$ & $4.3\pm0.01$ & & $-8.5\pm3.8$   & $4.0\pm0.01$ \\
\hline
   HEXTE    &   15--250\,keV   & $-2.3\pm1.8$ & $3.1\pm0.04$ & & $-11.0\pm16.5$ & $2.9\pm0.03$ \\
\hline
   LAT      & 0.1--300\,GeV    & $-0.8\pm0.7$ & $1.3\pm0.01$ & & $-3.1\pm6.3$   & $1.2\pm0.01$\\
\hline \hline
\end{tabular}
\end{center}
[a] These intercepts correspond to the values at MJD 55000.
\end{table}

\clearpage

\begin{table}
\footnotesize
\caption {The fitting results and correlation coefficients for the relationship
between $L_{X,\gamma}$ and $L_{sd}$}
\scriptsize{}
\label{table:2}
\medskip
\begin{center}
\begin{tabular}{c c c c c c c c c c c}

\hline \hline
Group &Instrument & Energy Band  & Parameter1 &  &  Parameter2 &  $m$  &   $\rm Log_{10}$ $n$  &  $\rho$  & $p$ \\
\hline
1 &{PCA} & 5--60\,keV & {$L_{X}$} & {vs.} & {$L_{sd}$}  &  $1.6\pm0.3$    &  $-26.1\pm10.4$  & $0.69$  & $<10^{-5}$  \\
\hline
2 &{HEXTE} & 15--250\,keV & {$L_{X}$} & {vs.} & {$L_{sd}$} &  $2.3\pm1.5$    &  $-54.4\pm56.1$  & $0.59$ & 0.05\\
\hline
3 &{LAT} & 0.1--300\,GeV & {$L_{\gamma}$} & {vs.} & {$L_{sd}$} &  $1.7\pm1.4$    &  $-29.7\pm54.1$  & $0.30$ & 0.28\\
\hline
4 & &  X-ray ($<$ 1\,MeV) &  $L_{X,th}^{\rm (OG)}$ & vs. & $L_{sd}^{\rm (OG)}$   &  $1.2$    &  $-9.5$        & ...  & ...  \\
\hline \hline
\end{tabular}
\end{center}
\end{table}

\end{document}